\documentstyle[12pt,aasms4,psfig]{article}

\newcommand{\xmm}{{\it XMM} }

\newcommand{\chandra}{{\it Chandra} }

\newcommand{\etal}{{\it et al.} }
\newcommand{\asca}{{\it ASCA} }

\newcommand{\fekalfa}{Fe-K }

\newcommand{\mcg}{MCG~$-$6$-$30$-$15 }

\newcommand{\figmcg}{Figure~1 }
\newcommand{\figsigew}{Figure~2 }
\newcommand{\figchisq}{Figure~3 }

\newcommand{\tabdblgauss}{Table~1 } 
\newcommand{\tartarus}{{\sc tartarus} }

\begin{document}

\title{
THE EFFECT OF CHANGES IN THE \asca CALIBRATION ON THE \fekalfa LINES IN 
ACTIVE GALAXIES}

\author{T. Yaqoob\altaffilmark{1,2},
U. Padmanabhan\altaffilmark{1},
T. Dotani\altaffilmark{3},
K. Nandra\altaffilmark{2,4}}

\begin{center}
{\it Accepted for Publication in the Astrophysical Journal, to appear in Vol. 569, 10 April 2002}

\end{center}

\altaffiltext{1}{Department of Physics and Astronomy,
Johns Hopkins University, 3400 N. Charles St., Baltimore, MD 21218, USA. }
\altaffiltext{2}{Laboratory for High Energy Astrophysics, 
NASA/Goddard Space Flight Center, Greenbelt, MD 20771, USA.}
\altaffiltext{3}{Institute of Space and Astronautical Science, 3-1-1, 
Yoshinodai, Sagamihara, Kanagawa 229-8510, Japan.}
\altaffiltext{4}{Universities Space Research Association.}

\begin{abstract}
The \asca calibration has evolved considerably since launch and indeed,
is still evolving. 
There have been concerns in the literature that changes
in the \asca calibration have resulted in the \fekalfa lines in active galaxies (AGN)
now being {\it systematically} narrower than was originally thought.
If this were true, a large body of \asca results would be impacted. 
In particular, it has been claimed that the broad red wing
(when present) of
the \fekalfa line
has been considerably weakened
by changes in the \asca calibration. 
We demonstrate 
explicitly that changes in the \asca calibration over 
a period of about eight years have a negligible effect
on the width, 
strength, or shape of the
\fekalfa lines.
The reduction in both width and equivalent width is only $\sim 8\%$
or less. We confirm this with simulations and individual sources, as well as sample
average profiles. The average profile for type 1 AGN is still very
broad, with the red wing extending down to $\sim 4$ keV. 
The reason for the claimed, apparently large, discrepancies
is that in some sources the \fekalfa line is complex, and
a single-Gaussian model, being an inadequate description
of the line profile, picks up different portions of the
profile with different calibration. 
Single-Gaussian fits do not therefore model all of the line
emission in some sources, in which case they do not compare old and current
calibration since the models do not then describe the data.
\end{abstract}

\section{INTRODUCTION}

The {\it Advanced Satellite for Cosmology and Astrophysics}
(\asca -- Tanaka, Inoue, \& Holt 1994) found that the  \fekalfa 
fluorescent emission line in Seyfert~1 galaxies is 
often very broad, 
and this is generally interpreted as the result of an origin in
matter in an accretion disk rotating around a central black hole
(see Fabian \etal 2000 and references therein).
The line profile is thought to be sculpted by characteristic
gravitational and Doppler energy shifts.
Currently, study of the  \fekalfa emission line is
the only way to probe
matter within a few to tens of gravitational radii
of a black hole. Indeed, the broad \fekalfa lines provide
some of the strongest evidence to date for the existence of
black holes.

It has been known since the early days of \asca that 
the \fekalfa line in AGN 
is not always tremendously broad (e.g. Ptak 1994) and that the
profile comes in a variety of
shapes (e.g. Yaqoob \& Weaver 1996; Nandra \etal 1997, hereafter N97). 
It has also been known
that Seyfert type 1.9--2.0 galaxies have a predominantly narrow
\fekalfa line, with FWHM less than $\sim 10,000$ km/s, probably
originating in cold matter far from the black hole (e.g. NGC 2992;
Weaver \etal 1996). It was apparent that many Seyfert 1.5--1.9 galaxies
likely have composite narrow and broad \fekalfa line components
(e.g. Weaver \etal 1993, Yaqoob \etal 1995, Weaver \etal 1997,
Weaver \& Reynolds, 1998).
\chandra and \xmm have now shown that composite narrow and
broad \fekalfa lines are common even in Seyfert 1 galaxies and
quasars (Yaqoob \etal 2001a, Kaspi \etal 2001, Reeves \etal 2000,
Pounds \etal 2001). 

Recently Lubi\'{n}ski and Zdziarski (2001; hereafter LZ01) have claimed that
the \fekalfa lines in AGN are {\it systematically} narrower and weaker 
(smaller equivalent width, or EW) than had previously been thought. 
They speculated that 
this result can be attributed to changes in the calibration of the \asca 
instruments. However they did not attempt to demonstrate this by
isolating calibration effects. 
The purpose of the present paper is to explicitly demonstrate that
changes in the \asca calibration have {\it not} resulted in any significant
changes in the width, equivalent width, or shape of the \fekalfa lines
in AGN. This issue is important to resolve because speculation 
about the \asca calibration has brought into question a significant 
body of \asca results and has rendered the 
astronomical community rather confused. We will also give the
real explanation of the results of LZ01, which turns out to
be a very subtle data analysis issue. 

\section{PRINCIPAL CALIBRATION CHANGES}
\label{asymline}

We shall compare the results of fitting various data sets with two
sets of calibration. The first, which we shall call `OLD', is the same,
or essentially the same,
as that used by N97. The second, which we shall call `CURRENT' is the
same as that used by LZ01.
The CURRENT calibration files are from the \tartarus 
AGN database\footnote{http://tartarus.gsfc.nasa.gov/}, except where noted.
All of the observations considered by LZ01 were made
before radiation damage made a significant impact on the
CCDs.
Therefore the principal differences
between the OLD and CURRENT calibration for these early data
are, (1) improvements in the SIS response matrix, (2)
multiplicative, energy-dependent corrections for the effective area
(the so-called `arf filter'), and (3)
changes in the X-ray Telescope (XRT) ray-tracing
(see Yaqoob \etal 2001b for further details on calibration history).
The latter is the {\it only} change that can make the
OLD and CURRENT calibration have different effects
on different observations since the first two are
independent of position on the detector and independent
of time of the observation
(as long as that time is early enough in the mission).
Even so, the XRT response is a
smooth function of source off-axis angle, and its variation
is small from observation to observation,
given the limited range of off-axis angles
in observations. 

We performed simulations of SIS0$+$SIS1
spectra in order to isolate calibration effects
from data processing differences and
statistical effects (the latter by using a very large exposure time 
of $10^{10}$~s).
We used the original (OLD) calibration files used by Tanaka \etal (1995)
in their analysis of \asca AO2 data for MCG$-$6$-$30$-$15.
First, we used one of the best-fitting relativistic
Schwarzschild disk-line models of Tanaka \etal (1995), namely,
the model with the \fekalfa rest energy fixed at 6.4 keV,
continuum power-law index of 1.96,  and
the radial line-emissivity power-law index, $q$, fixed at $-3$.
We then investigated the effect of using CURRENT 
calibration (from \tartarus) with the
spectra that were simulated using the OLD calibration. 
The line profiles, in the form of ratios of data to
power-law models fitted over the 3--4 keV and 8--10 keV bands 
for each calibration, are compared in \figmcg~(a). Indeed, it can be
seen that the CURRENT calibration reduces the red wing of the line.
However, spectral fitting of the simulated data with the CURRENT
calibration shows that the {\it EW is reduced by only 8\%}.
Note that residuals from the latter fit are never more than 3\%.
We repeated this excercise with a Gaussian line profile, using
parameters which are typical of Seyfert type 1 galaxies with some
of the broadest lines. Using the same continuum as above,
we simulated
a Gaussian with $\sigma = 0.6$ keV, $\rm EW=300$ eV, centered at
6.4 keV. The line profiles for the OLD and CURRENT calibration are
compared in \figmcg~(b). The red side is slightly diminished and the
blue side slightly enhanced for the CURRENT calibration relative
to OLD. However, spectral fitting reveals that the {\it width 
($\sigma$) is reduced only  by
8\%, and the EW only by 3.6\%}. There is an apparent slight shift in the
centroid energy but this is only by 2\% and will also depend on the
actual width of the line. {\it It is important to note that this
apparent shift in the centroid energy is NOT due to changes in the
instrumental energy scale, which have been measured to be much less than
this.} Rather, the apparent shift is due to different weights given
to the red and blue sides of the broad Gaussian by the changes
in effective area.

These results are of course confirmed for the real data from
the \asca AO2 observation of MCG$-$6$-$30$-$15. \figmcg(c)
shows the Fe-K line profiles (produced in the same way as
for the simulated data) for the OLD and CURRENT calibration.
Now statistical effects make it very hard to see any difference
in the profiles.
We fitted the real 3--10 keV SIS0$+$SIS1 data with a power-law plus relativistic
Schwarzschild disk-line model.
The only parameters that were fixed were $q=-3$, and the rest
energy of the line (6.4 keV).
We obtained the following (Tanaka \etal 1995 values in parentheses):
a disk
inclination angle of $31.6^{+2.1}_{-2.2}$ 
($29.7^{+2.9}_{-3.9}$) degrees,
an inner disk radius of $7.5^{+2.5}_{-1.1}$ ($7.4^{+3.6}_{-1.4}$)
$r_{g}$, an outer disk radius of
$20.9^{+28.5}_{-6.3}$ ($20.0^{+25.0}_{-6.4}$) $r_{g}$, and
an EW of $344^{+79}_{-74}$ ($380^{+100}_{-110}$) eV.
Yaqoob etal (2001b) also show that a double-Gaussian model
for the \fekalfa line again gives
completely consistent results for the Tanaka \etal (1995) and
\tartarus data and calibration.

\section{THE N97  SAMPLE}
\label{n97sample}

In the previous section we explained why changes in \asca calibration
should not have a significant impact on the \fekalfa line
parameters for any of the observations considered by LZ01.
Nevertheless, we attempted to discover the origin of the
results of LZ01. For this we concentrated on the N97
sub-sample, since these AGN have some of the broadest lines.
N97 could not constrain the
line width for all observations in their sample and,
excluding these, as well as NGC~4151 and NGC~6814
(LZ01 omitted these latter two),
we have used fourteen observations of twelve sources
to directly compare with N97. These AGN are 
F9 (AO1), 3C~120 (AO1), NGC~3227 (AO1), NGC~3516 (AO1), Mkn 766 (AO1), 
IC~4329A (PV), NGC~5548 (PV),
Mkn 841 (PV, sequence 70009000), Mkn~509 (AO1), 
NGC~7469 (AO1, sequence 71028030), \mcg (PV), NGC~3783 (AO1).
The latter two sources have two observations each in this sample
and we use the same notation, (1) and (2), as in N97 to distinguish
them.
Note that `PV' refers to the `Performance Verification' phase
of {\it ASCA}. 

We then repeated the 
N97 3--10 keV power-law plus Gaussian spectral fits using the CURRENT
calibration by
using the
actual spectral and calibration files used by LZ01
(i.e. from {\sc tartarus}), except for two sources
(F9, and NGC~3227) which were not available in the
{\sc tartarus} database at the time. For these, we reduced
the data following the methods described in Yaqoob \etal (1998).
As will become clear, our conclusions do not depend on these
details of data reduction. For convenience
we shall take the liberty of referring to all fourteen of these datasets
as {\sc tartarus} spectra and calibration.
We used four instruments (SIS,GIS) fitted simultaneously,
as did N97 and LZ01. 
The best-fitting line widths and their mean statistical errors
obtained using the CURRENT calibration are plotted against the
corresponding N97 values (which are based on 
the OLD calibration) in
Figures~2a and 2b respectively. 
The corresponding equivalent widths (EW) and
mean statistical errors are shown in Figures~2c and 2d.
The
statistical error shown here is the straight mean of
the positive and negative error on each 
measurement, for
a $\Delta \chi^{2}$ criterion of 4.7 (this exact value was used
by N97 and corresponds to 68\% confidence for three
parameters). 

\figsigew shows that all the results from {\sc tartarus} and
N97 are consistent except for three observations.
These are NGC~3783(1), Mkn841, and NGC~3516, for which the 
{\sc tartarus} spectral
fits indicate significantly narrower \fekalfa lines than N97. 
In order to investigate how such apparently huge 
differences in the best-fitting parameters
could come about, for these sources we 
emulated OLD calibration 
for the {\sc tartarus} spectra in order to isolate
calibration effects and separate them from
data reduction and statistical effects.
To do this we used old SIS response matrices, as in
N97, ({\tt s0c1g0234p40e1\_512v0\_8i.rmf}, 
and {\tt s1c3g0234p40e1\_512v0\_8i.rmf}), 
disabling the 
multiplicative effective area factors in {\tt ascaarf},
and using v1.1 of the XRT response (see Yaqoob \etal 2001b 
for details on all these files and procedures).
First, we compared the SIS \fekalfa line profiles with this
OLD and CURRENT calibration by fitting a power-law
continuum to the 3-4 keV and 8-10 keV band only and
computing the ratios of the data to this continuum model.
Note that OLD and CURRENT calibration fits were  
performed separately, since that is how they would be
compared in the literature.
\figmcg~(d)--(f) shows the results, namely that the {\it line profiles
for all three sources are very broad and the difference
between OLD and CURRENT calibration is negligible}.
Yet the formal best-fits from spectral fitting yield
disparate results: according to LZ01, these three very broad lines
are some of the narrowest. 
Table~1 shows the single-Gaussian results for each of these
three sources
and each calibration. We are able to reproduce the
results of N97 with the OLD calibration and the results
of LZ01 with the CURRENT calibration.
How can the broad line profiles
in \figmcg~(d)--(f) yield such narrow line widths from
fits with the CURRENT calibration?
One gets a clue
by noticing that, in each source,
single-Gaussian models of the line have two minima which
are very close in $\chi^{2}$ space (see Figure~3). In fact,
if one correctly accounts for the double minima in
$\chi^{2}$ space, the differences in the widths from
the single-Gaussian fits are not statistically significant (see Table~1).
Also, it is clear from
\figmcg~(d)--(f) that the narrow-line solutions are only modeling
the core of a very broad profile. Clearly, 
{\it regardless of the calibration}, a single Gaussian
is inadequate to model these line profiles, which are complex.
However, since the continuum slope is slightly different
for the OLD and CURRENT calibration, the formal best-fit
can be nudged from modeling only the broad part or only
the narrow core of the line.
If we fit these complex lines with a double-Gaussian model
we get completely consistent results between the OLD
and CURRENT calibration (see Table~1). Moreover, the addition
of the second Gaussian is highly statistically significant
for each source and each calibration
($\Delta \chi^{2}$ ranges between 13.7--36.5,
for the addition of two free parameters; note that the width of
the narrow Gaussian is frozen at 1 eV in these fits since the
width is consistent with zero).
An important lesson here is that {\it it is erroneous to make statements
about instrument calibration, and indeed to make astrophysical inferences,
based on a model which does
not describe the data}.
Single-Gaussian spectral fits which only model the narrow core of the line
seriously underestimate the line width and equivalent width.
Moreover, attempts to relate measurements of the narrow core
to relativistic disk models of the line are erroneous since
the core of the line very likely does not come from the disk,
and the part that does, is not measured with the single-Gaussian,
narrow-line fits using the CURRENT calibration.
The results of N97, which modeled the
broad component in the case of complex lines,
are actually a fairer respresentation of the overall
width and total equivalent width of the \fekalfa lines.

\section{COMPOSITE LINE PROFILES}

We have also constructed and compared the SIS0$+$SIS1 composite data/model ratios for
two samples, both drawn from N97. The first sample excludes only
the observations of NGC~6814, MCG~$-$6$-$30$-$15, and NGC~4151 from N97 
(leaving seventeen observations).
The second sample is comprised of the same fourteen observations which were used
in the spectral analysis described above and shown in Figure~2. 
In all the following, Fe-K profiles were constructed by computing ratios
of data to power-law models fitted to individual
observations over the 3--4 keV and 8-10 keV bands. 
Figure~4(a) shows the composite profile for sample~1, using the original data
files and calibration from N97 (crosses), with the corresponding composite
using \tartarus data files and calibration (filled circles).
Figure~4(b) shows the composite profile for sample~2, this time {\it using only
the} \tartarus {\it spectral files} but first fitted with emulated OLD calibration
(crosses), and then compared to fitting with the
\tartarus (CURRENT) calibration. The OLD calibration was
emulated as described above in \S 2 and 3.
Three things are abundantly clear from Figure~4.
(1) Calibration differences (and for that matter, data extraction differences)
do not have a discernable effect on the composite profiles; (2) all the composites
are very broad, with some asymmetry in the red wing, which extends down to
$\sim 4$ keV; (3) inclusion of  MCG~$-$6$-$30$-$15 in the composite does not
bias the composite profiles. 

LZ01 found an apparently large difference in the
comparison of composite line
profiles from N97 and \tartarus (see their Figure~2). This difference was
exaggerated due to at least two factors. First, they compared a
scaled version of the fit to the N97 composite line profile with their own
composite data/model ratio. The line profile and data/model ratio are
not equivalent, with the former being the latter multipled by the
continuum slope. As this is steep, the effect is to de-emphasize the
strength of the red wing (and enhance the blue) in the appearance of
the ratio. LZ01 used this to support their scientific conclusion, when
it is in fact a difference in presentation of the data. 
Moreover, LZ01 used a slightly different
method for calculating the average ratio compared to N97.
Before averaging data/model ratios from different sources
they must first be rebinned so that they all have the same binning.
In this binning process, LZ01 attempted to
account for the fact that some input bins overlap more than one output
bin since the input bin widths could be larger than the output bin widths. 
However, they
distribute
the value in an input bin
over the output bins by an amount proportional to the
ratio of the width of the output bin to the input bin width.
In the case that an input bin boundary falls inside an output bin,
it is the fraction of the output bin width that overlaps the input bin
that is used.
LZ01 explicitly used this method to conserve numbers of photons
during the rebinning.
However, this method is incorrect
since we are dealing with a {\it ratio}.
Only the error in the ratio should be weighted by the
method described above,
not the ratio itself. This error in
LZ01 causes them to
systematically underestimate the data/model ratio in all bins
because the output bins will have, most of the time, less than the
true data/model ratio, and never more. Also, since the input bin-width
distribution could be very different for the different data sets
that the average is comprised of, it is
even possible that the average ratio is distorted.
These errors in LZ01 are the cause of the discrepancies with N97.
The ratios shown in
our Figure~4 were computed by correct 
rebinning of the input data/model ratios but we note that this
causes only very small differences in the ratio compared to the
method of N97 (in which fractional bin overlaps
were not treated rigorously).

We reiterate here something that has
already been stated in N97. That is,
mean line profiles serve only an illustrative purpose
and it can be misleading
to try and model them since they are obtained by adding 
sources with different continuum spectra, line shapes, and fluxes.
For example, suppose we have a sample in which 50\% of the
line profiles have a strong red wing and the other 50\% have
a strong blue wing. From the composite spectrum one would conclude
that the average source has a symmetric line profile, when in
fact not a single source may possess such a profile.
Clearly, analyses which model average profiles presuppose that
the individual profiles and continuua are similar, but this
need not be true at all. 

\section{CONCLUSIONS}
\label{concl}

We find that the width, equivalent width, and shape of
the \fekalfa lines in AGN have not been significantly
affected by changes
in the \asca calibration over a period of more than eight years.
In particular, the width and equivalent width change by only
$\sim 8\%$ or less. The red wing, in the case of a highly
skew \fekalfa line profile as in \mcg is reduced by a
similar magnitude. In most cases the statistical errors
are larger, so the changes in \asca 
calibration are inconsequential for
almost all \asca observations. We also compared the
sample composite \fekalfa line profiles using 
OLD and CURRENT calibration and find that the differences
are still not statistically significant.
In some individual cases the \fekalfa lines are complex so single-Gaussian
models are inadequate for describing the profile and
consequently only model a portion of the line.
Then, the line width and equivalent width may
be seriously understimated, leading to erroneous
conclusions about the astrophysics of the
line and about instrument calibration. 
When adequately modeled, we again find no significant difference in the
\fekalfa line parameters with respect to changes in calibration.

The authors acknowledge support from NASA grants NCC-5447 (TY, UP), 
NAG5-10769 (TY), and NAG5-7067 (KN).
This research made use of the HEASARC online data archive
services, supported by NASA/GSFC. It also made use of
the {\sc tartarus} AGN database which is supported by
Jane Turner and Kirpal Nandra under NASA grants NAG5-7385 and NAG5-7067.
Special thanks to Prof. Y. Tanaka for his inputs and motivation.
The authors also wish to thank Koji Mukai, Ken Ebisawa, Keith Arnaud, and
other members of the ASCATEAM for their input to this work.
They also thank
Jane Turner, Ian George, Kim Weaver 
for helpful discussions and comments.
We are grateful to Andrzej Zdziarski, the referee, and Piotr
Lubi\'{n}ski, for making suggestions to greatly improve this paper.

\newpage

\begin{deluxetable}{lcccccccc}
\tablecaption{Single and Double-Gaussian Models for NGC 3783, Mkn 841, and NGC 3516}
\tablecolumns{9}
\tablewidth{0pt}
\tablehead{ 
\colhead{} & \colhead{$E_{N}$} & \colhead{$\sigma_{N}$} & \colhead{$EW_{N}$} & \colhead{$E_{B}$} & \colhead{$\sigma_{B}$} & \colhead{$EW_{B}$} & \colhead
{$\chi^{2}$(d.o.f)} & \colhead{$\Delta \chi^{2 \ \dagger}$} \nl 
\colhead{} & \colhead{(keV)} & \colhead{(keV)} & \colhead{(eV)} &
\colhead{(keV)} & \colhead{(keV)} & \colhead{(eV)} & \colhead{ } & \colhead{} \nl
}

\startdata

NGC3783(1) & & & & & & & & \nl
S (OLD)  & $^{}_{}$ & $^{}_{}$ & $^{}_{}$ & $6.13^{+0.16}_{-0
.19}$ & $0.69^{+0.36}_{-0.27}$ & $527^{+346}_{-219}$ & $804.5(7
84)$ & $ $ \nl

S (CURRENT)  & $6.41^{+0.16}_{-0.50}$ & $0.0^{+1.09}_{-0.00}$ 
& $169^{+45}_{-40}$ &$^{}_{}$ & $^{}_{}$ & $^{}_{}$ & $806.7(7
84)$ & $ $ \nl

D (OLD)  & $6.41^{+0.05}_{-0.06}$ & $0.001^{}_{}(f)$ 
& $114^{+44}_{-59}$ & $5.88^{+0.32}_{-0.41}$ & $0.91^{+0.91}_{-0
.44}$ & $469^{+761}_{-253}$ & $777.5(782)$ &$27.0$ \nl

D (CURRENT)   & $6.41^{+0.54}_{-0.55}$ & $0.001^{}_{}(f)$     
& $129^{+54}_{-54}$ & $5.88^{+0.36}_{-0.44}$ & $1.35^{+0.71}_{-0
.30}$ & $700^{+474}_{-314}$ & $777.3(782)$ & $29.4$ \nl

Mkn 841 & & & & & & & & \nl
S (OLD)   & $^{}_{}$ & $^{}_{}$ & $^{}_{}$ & $6.11^{+0.41}_{-0
.26}$ & $0.70^{+0.35}_{-0.66}$ & $738^{+575}_{-355}$ & $422.6(3
98)$ & $ $ \nl

S (CURRENT)   & $6.48^{+0.09}_{-0.08}$ & $0.09^{+0.79}_{-0.09}$ 
& $257^{+116}_{-101}$ & $^{}_{}$ & $^{}_{}$ & $^{}_{}$ & $419.8(
398)$ & $ $ \nl

D (OLD)   & $6.51^{+0.10}_{-0.12}$ & $0.001^{}_{}(f)$     
& $154^{+128}_{-102}$ & $5.87^{+0.37}_{-0.49}$ & $0.72^{+0.56}_{
-0.48}$ & $556^{+672}_{-288}$ & $407.2(396)$ & $15.4$ \nl

D (CURRENT)   & $6.51^{+0.11}_{-0.09}$ & $0.001^{}_{}(f)$ 
& $250^{+96}_{-178}$ & $5.64^{+0.67}_{-0.42}$ & $0.59^{+1.09}_{-
0.30}$ & $324^{+885}_{-231}$ & $406.1(396)$ & $13.7$ \nl

NGC3516 & & & & & & & & \nl
S (OLD)   & $^{}_{}$ & $^{}_{}$ & $^{}_{}$ & $6.09^{+0.12}_{-0
.14}$ & $0.65^{+0.22}_{-0.19}$ & $360^{+143}_{-106}$ & $873.4(7
70)$ & $ $ \nl

S (CURRENT)   & $6.33^{+0.07}_{-0.22}$ & $0.23^{+0.60}_{-0.10}$
& $171^{+245}_{-45}$ & $^{}_{}$ & $^{}_{}$ & $^{}_{}$ & $884.3(7
70)$ & $ $ \nl

D (OLD)   & $6.41^{+0.04}_{-0.05}$ & $0.001^{}_{}(f)$
& $75^{+22}_{-40}$ & $5.86^{+0.36}_{-0.22}$ & $0.77^{+0.38}_{-0
.11}$ & $294^{+233}_{-84}$ & $840.6(768)$ & $32.8$ \nl

D (CURRENT)   & $6.41^{+0.08}_{-0.04}$ & $0.001^{}_{}(f)$
& $79^{+30}_{-34}$ & $5.91^{+0.37}_{-0.33}$ & $1.00^{+0.93}_{-0
.24}$ & $318^{+581}_{-115}$ & $847.8(768)$ & $36.5$ \nl

\enddata
\vspace{-5mm}
\tablecomments{ 
Comparison of the OLD and CURRENT calibration for single (S) and 
double-Gaussian (D) models of the \fekalfa line.
Subscripts $N$ and $B$ denote the parameters 
(center energy,  $E$, intrinsic width, $\sigma$, and equivalent
width, $EW$) for the narrow and broad \fekalfa line components respectively.
Statistical errors are 68\% confidence for four
($\Delta \chi^{2} = 4.70$) and six ($\Delta \chi^{2} = 7.00$)
interesting parameters, for the single and double-Gaussian
models respectively. 
$^{(\dagger)}$ $\Delta \chi^{2}$ is the difference in
$\chi^{2}$ between single and double-Gaussian fits.
}
\end{deluxetable}

\newpage
\section*{Figure Captions}

\par\noindent
{\bf Figure 1} \\
(a) Isolation, from other effects, of differences between
OLD and CURRENT \asca calibration and their effect on 
a typical \fekalfa relativistic disk line
profile. Shown is a simulation of co-added SIS0 plus SIS1 data
for the relativistic disk line model for
\mcg of Tanaka \etal
(1995) described in the text, using OLD calibration
(solid line). An extremely large exposure time was used ($10^{10} \ s$) to
eliminate statistical effects. The dotted line shows the
line profile one would deduce if
the simulated data were then fitted in the same
way as real data, using the CURRENT calibration instead. 
(b) As (a) but for a typical broad Gaussian line profile
with parameters as described in the text.
(c)
Effect of changes in the \asca calibration on
the \fekalfa line profile using the actual data for \mcg (\asca AO2).
SIS0 and SIS1 are data co-added. See text for details.
Crosses correspond to the original Tanaka \etal (1995) data
and calibration, and 
filled circles correspond to \tartarus data and calibration.
(d), (e), (f) 
The SIS0+SIS1 \fekalfa line profiles for the three data
sets 
which show the largest discrepancies 
in
spectral
fitting results
between the OLD and CURRENT calibration. Yet the
line profiles using the OLD and CURRENT calibration (crosses and filled circles
respectively) are virtually identical. All three are {\it very broad}
and have large equivalent widths (see \tabdblgauss). The line
profiles are complex and
cannot be adequately modeled with a single Gaussian.
Thus, a single Gaussian only models the core of the
line if the CURRENT calibration is used.
Consistent results are obtained between OLD and CURRENT calibration
when a double-Gauissian model is used (see \tabdblgauss).

\par\noindent
{\bf Figure 2} \\
Comparison of the N97 
and \tartarus results
for the deduced width and equivalent width
of the \fekalfa lines for 14 observations of 12 AGN,
when modeled only with a single Gaussian
(top panels). Also shown are the corresponding
mean statistical errors of the measurements (bottom panels).
The observations are from  
a sub-sample which is common to both the samples of
N97 and LZ01 (see text).
The solid lines correspond to the 
case if the N97 data (OLD calibration)
and \tartarus data (CURRENT calibration)
were to give identical results. The statistical
errors are
the straight means of the positive and negative errors
deduced from a $\Delta \chi^{2} = 4.7$ criterion, identical
to that used by N97, and corresponds to 68\% confidence for
three interesting parameters.
Filled circles are the direct comparisons for single-Gaussian fits.
Open circles,
for NGC~3783(1), NGC~3516, Mkn~841,
correspond to the second minimum
from the single-Gaussian fits with \tartarus data (also see \figchisq).
\tabdblgauss shows that the broad-line widths and equivalent widths
using double-Gaussian models are completely consistent for
the OLD and CURRENT calibration and correspond to these
second minima.
The single-Gaussian
narrow-line solutions (indicated by arrows)
obviously understimate the true
width and equivalent width of the lines much more than
the broad-line solutions. The 
actual line profiles for these three cases are shown in \figmcg
and are {\it very broad}.

\par\noindent
{\bf Figure 3}
The change in $\chi^{2}$ as the line width in the single-Gaussian model
is varied away from the best-fit with the CURRENT calibration for
NGC~3783(1). It can be seen that there are two minima, which correspond
to the solutions of LZ01, N97. This indicates that the line is complex
and that a single Gaussian is inadequate to model the line, which is
very broad (see \figmcg).
 
\par\noindent
{\bf Figure 4}
Average SIS0$+$SIS1 Fe-K line profiles in the form ratios of data to power-law
models fitted independently to individual observations. (a)
Seventeen observations from
the N97 sample (excluded are observations of MCG~$-$6$-$30$-$15, NGC~4151 and NGC~6814).
Crosses are the original N97 data and calibration. Filled circles
are the \tartarus data and calibration. (b) Isolation of calibration
effects for the N97 sub-sample of fourteen observations discussed in this
paper. Crosses correspond to emulated OLD calibration, and filled
circles correspond to the \tartarus (CURRENT) calibration. In both cases
the \tartarus spectra were used, thus eliminating effects of data extraction
and processing.  
\newpage

\begin{figure}[h]
\vspace{10pt}
\centerline{\psfig{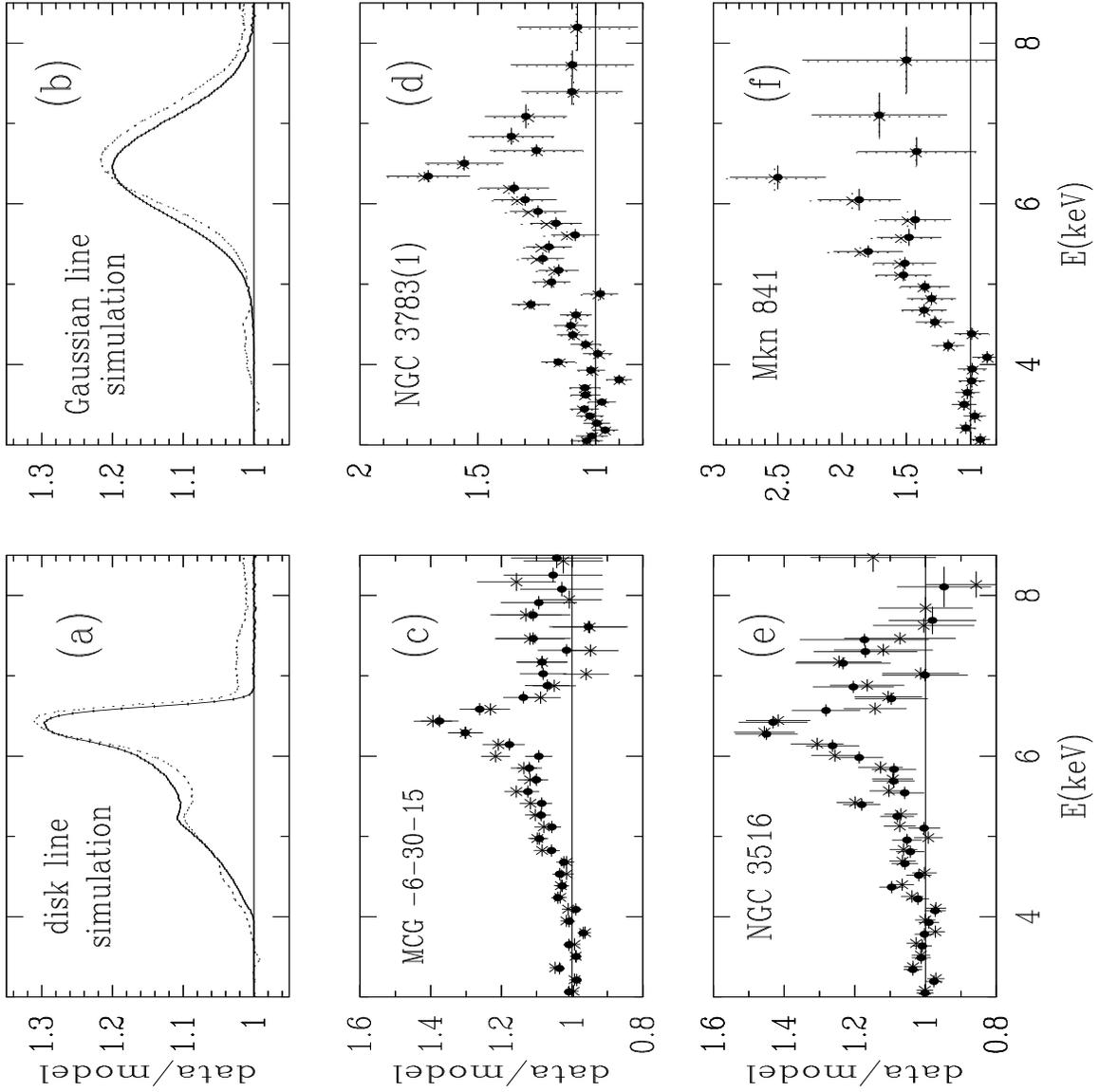}
}
\caption{See figure caption.}
\end{figure}
 
\begin{figure}[h]
\vspace{10pt}
\centerline{\psfig{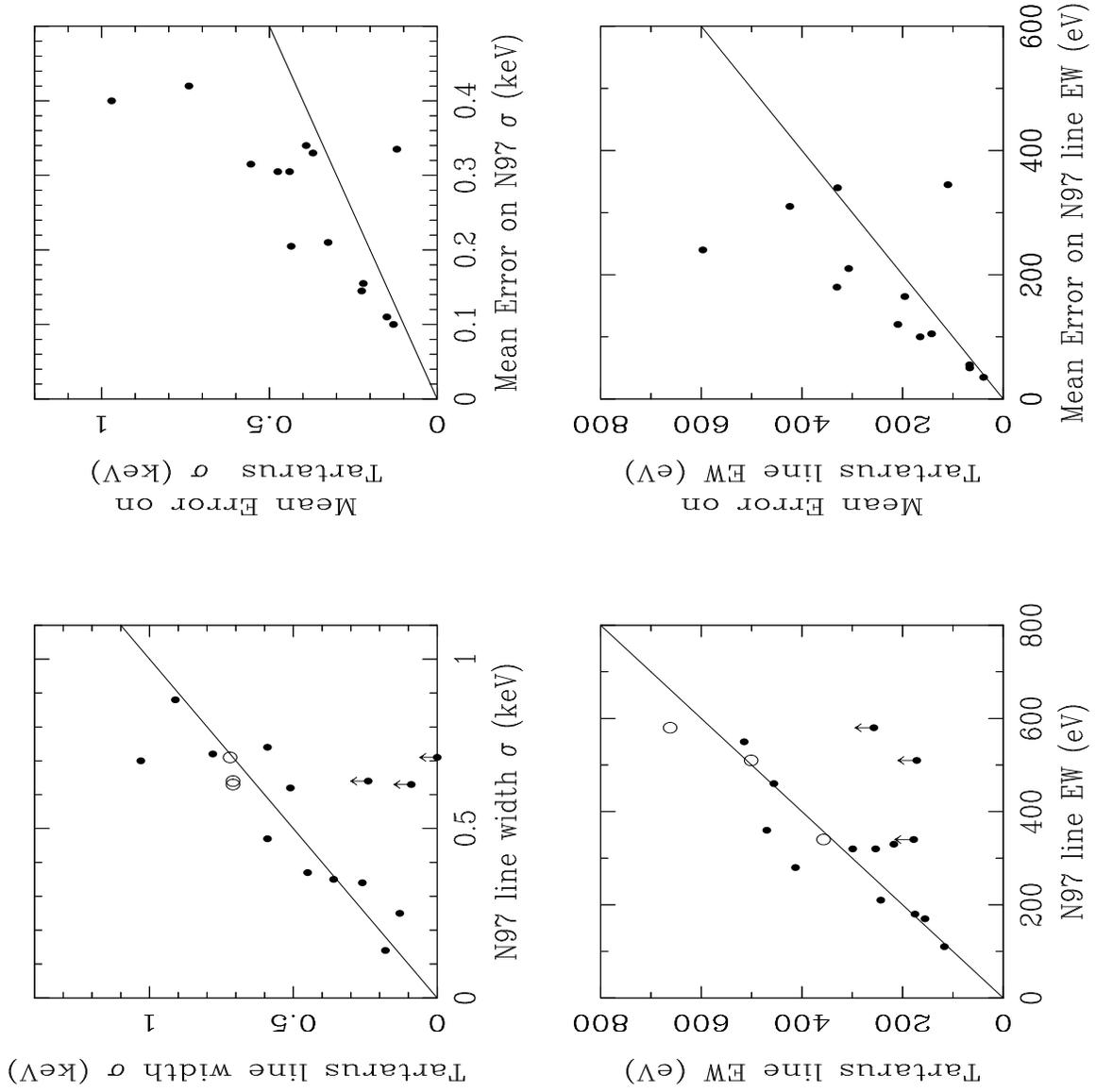}
}
\caption{See figure caption. }
\end{figure}

\begin{figure}[h]
\vspace{10pt}
\centerline{\psfig{file=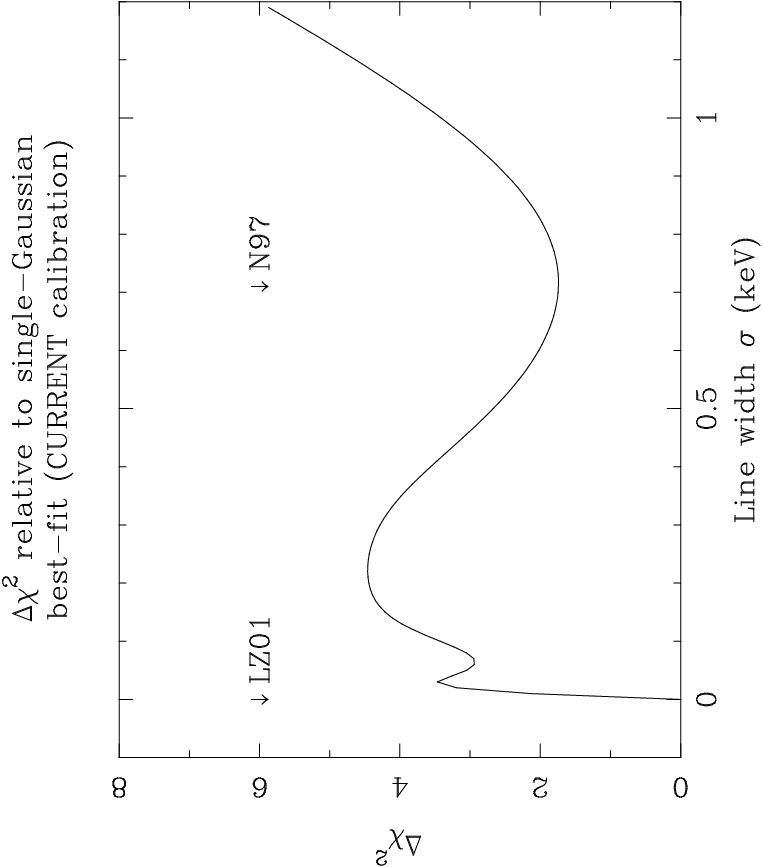,width=6.0in,height=6.0in}
}
\caption{See figure caption. }
\end{figure}

\begin{figure}[h]
\vspace{10pt}
\centerline{\psfig{file=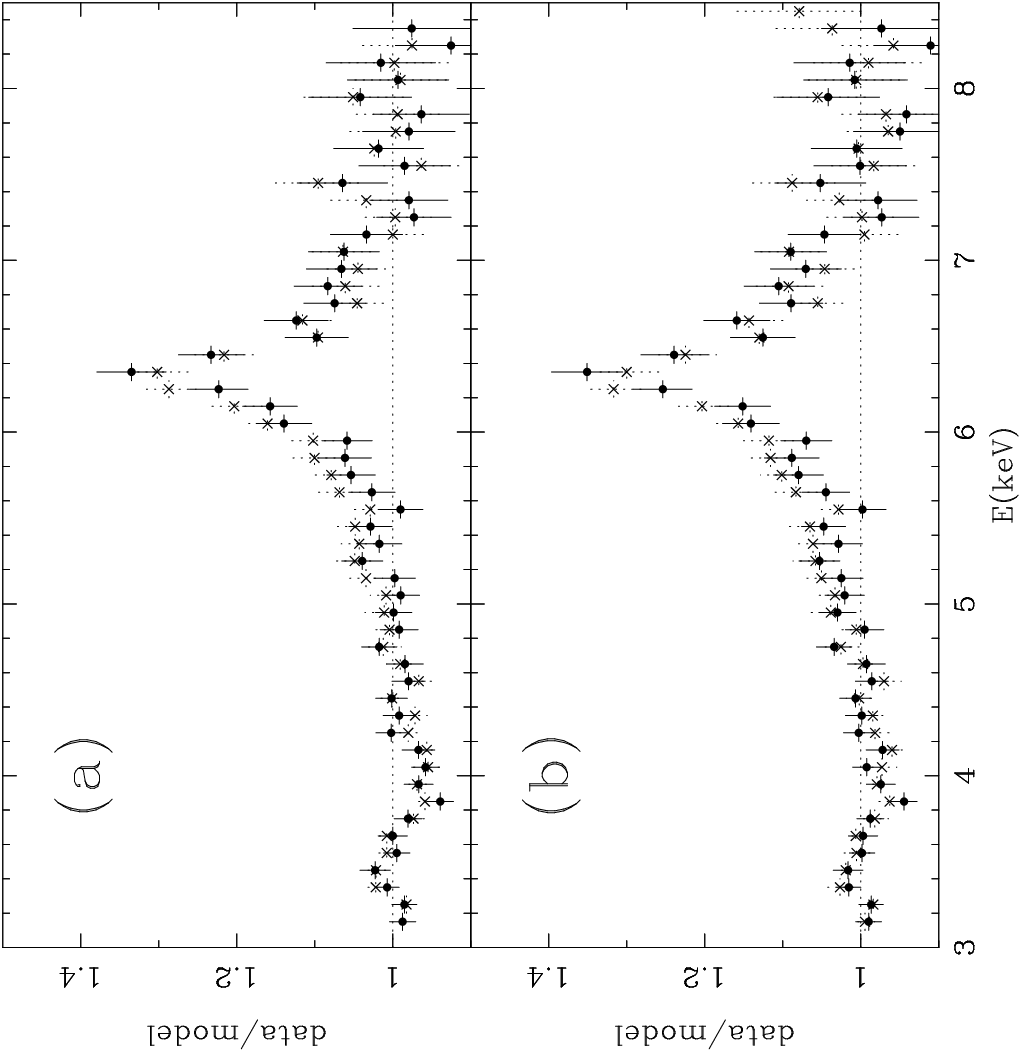,width=6.0in,height=6.0in}
}
\caption{See figure caption. }
\end{figure}

\end{document}